\documentclass{PoS}

\title{Spectator charge splitting of directed flow
              in heavy ion collisions}

\ShortTitle{Directed flow}

\author{\speaker{Antoni Szczurek}$^{a,b}$ and Andrzej Rybicki$^{a}$
\thanks{AS thanks organizers of EPS2013 conference 
for very good organization and atmosphere.}\\
        \llap{$^a$}Institute of Nuclear Physics, PL-31-342 Krak\'ow\\
        \llap{$^b$}University of Rzesz\'ow, PL-35-959 Rzesz\'ow\\
        E-mail: \email{antoni.szczurek@ifj.edu.pl, andrzej.rybicki@ifj.edu.pl}}

\abstract{We estimate the effect of the spectator charge on distortion
  of single charged pion distributions as well as on azimuthal
  anisotropies in heavy ion collisions. A large electromagnetic 
  effect on directed flow $v_1$ is
  predicted in good agreement with existing WA98 as well as RHIC data. 
  This effect
  results in a splitting of $v_1$ for positive and negative pions.
  Detailed analysis of this phenomenon may provide new
  information on the collision
  dynamics.}

\FullConference{The European Physical Society Conference on High Energy Physics -EPS-HEP2013\\
		18-24 July 2013\\
		Stockholm, Sweden}

\begin{document}

\section{Introduction}

{\em Noncentral nucleus-nucleus collisions} unambigously lead to 
azimuthal asymmetries and presence of spectators.
{\em Azimuthal correlations} between particles and the reaction plane
are one of the main subjects of study in heavy ion collisions.
They provide information about {collective effects}. On the other hand,
the presence of {\em charged fast moving spectators} generates 
strong electromagnetic fields.
The resulting {electromagnetic effects} {\em modify single 
particle spectra} \cite{RS2007}.
Recently, we discussed how these electromagnetic effects {\em influence 
also azimuthal correlations} \cite{RS2013}.

Our first constatation \cite{RS2007} that the EM fields generated by 
the remnants (spectators) of peripheral collisions distort the charged 
pion spectra was supported by precise NA49 experimental data
at $\sqrt{s_{NN}}$ = 17.3 GeV \cite{Rybicki2011}.
Spectacular effects were predicted and observed:
      \begin{itemize}
        \item A dip in $\pi^+$ density at pion $x_F$ = 0.15;
        \item An accumulation of strength for $\pi^-$ at $x_F$ = 0.15.
      \end{itemize}

In this presentation we first remind the effect of distortions of
single particle spectra, and then discuss the influence of electromagnetic
fields generated by the spectators on {\em directed flow}. 
Results of theoretical
calculations are compared to experimental data \cite{wa98,STAR_data}.
 The results included in this presentation are partially based on our precedent
 works \cite{RS2007,RS2013}.

\section{Sketch of the model}

\begin{figure}[htb] 
\begin{center}
\includegraphics[width=8cm]{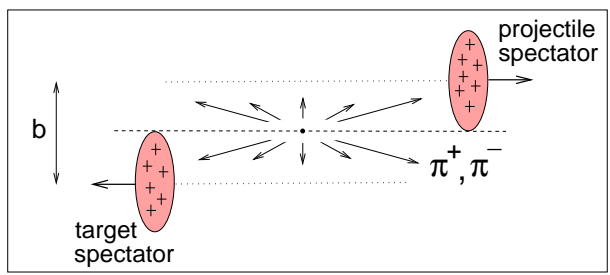}
\caption{\it Our simplified view of a Pb+Pb collision. The hypothetical
pion emission region is reduced to a single point in position space. 
\label{ideowy}
}
\end{center}
\end{figure}

In Fig.1 we present a general view of our model of the peripheral collision.
The situation can be summarized as follows:

 \begin{itemize}
 \vspace{-0.1cm}
 \item
 the collision takes place at a given impact parameter $b$.
 \vspace{-0.1cm}
 \item
 the two charged spectator systems follow their initial path. 
 \vspace{-0.1cm}
 \item
 the participating system evolves until pions are produced.
 \vspace{-0.1cm}
 \item
 charged pion trajectories are modified by the EM interaction.
 \vspace{-0.1cm}
 \item
 the spectator systems undergo a complicated, not fully understood, 
 nuclear deexcitation/fragmentation process.
 \end{itemize}

The initially produced charged pions are subjected to 
the EM field of the two spectator systems.
The spectator velocity  remains constant and identical 
to the velocity of the parent Pb ion.
We choose the overall CM system to calculate
the evolution of pion trajectories. We note that 
for symmetric Pb+Pb collisions,
this is also the N+N CM system.

The azimuthal correlations are usually quantified in terms
of the Fourier coefficients of the azimuthal distribution 
of the outgoing particles with respect to the reaction plane:
\begin{equation}
v_n \equiv <cos[n(\phi - \Psi_r)]>,
\end{equation}
where $\phi$ is the azimuthal angle of the emitted particle (pion),
while $\Psi_r$ is the orientation of the reaction plane defined
by the impact parameter vector $\vec{b}$.

The first order coefficient 
\begin{equation}
v_1 \equiv <cos(\phi - \Psi_r)>,
\end{equation}
 reflects the sideward collective motion and is known as directed flow. 
There exist rich data on $v_1$ from FOPI, E877, WA98, NA49, STAR, ALICE 
and other experiments. With some exceptions which will be discussed in 
this paper, these data are obtained for both charges of pions.

What is known about directed flow?
The underlying symmetry imposes its dependence on particle rapidity as an
asymmetric function ($v_1(y) = -v_1(-y)$).
The Glauber approach gives tilted initial conditions which
leads to tilted pressure and hydrodynamics produces final $v_1$ \cite{BW}.
The effect drops with collision energy.

The effects of the EM interaction discussed in this paper 
($v_1^{\pi,EM}$) will add up to these imposed by the strong force 
($v_1^{flow}$). For the final state values of $v_1$ 
observed in experiment, we expect as a first approximation:
\begin{center}
 \vspace{-0.4cm}
$v_1^{\pi^+} \approx v_1^{flow} + v_1^{\pi^+,EM}$ , \\
$v_1^{\pi^-} \approx v_1^{flow} + v_1^{\pi^-,EM}$ .
\end{center}
 
For simplicity, in this presentation we shall discuss only the pure 
electromagnetic component of directed flow. We note that this 
simplification does not affect the generality of the conclusions 
presented in this paper.

\section{Distortions of inclusive spectra of charged pions}

\begin{figure}         
\begin{center}
\includegraphics[width=4.7cm]{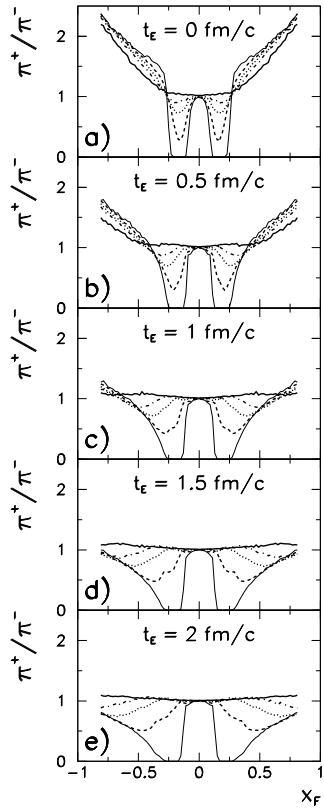}
\end{center}
 \vspace{-0.2cm}
\caption{Ratio of density of produced $\pi^+$ over produced $\pi^-$ in 
the final state of the peripheral Pb+Pb reaction, obtained for five values 
of the pion emission time $t_E$. The $\pi^+/\pi^-$ 
ratio is drawn as a function of $x_F$ at $p_T=25$ MeV/c (thin solid), 75 
MeV/c (dash), 125 MeV/c (dot), 175 MeV/c (dash-dot), and 325 
MeV/c (thick solid). 
}
\label{fig:ratio1}
\end{figure}

First we wish to remind how the electromagnetic effects distort
inclusive spectra of pions.
In Fig.\ref{fig:ratio1} we show the ratio of cross sections for $\pi^+$
and $\pi^-$ production as a function of pion $x_F$ for different
pion emission times defined in Ref.\cite{RS2007}.

\begin{figure}         
\begin{center}
\includegraphics[width=6cm]{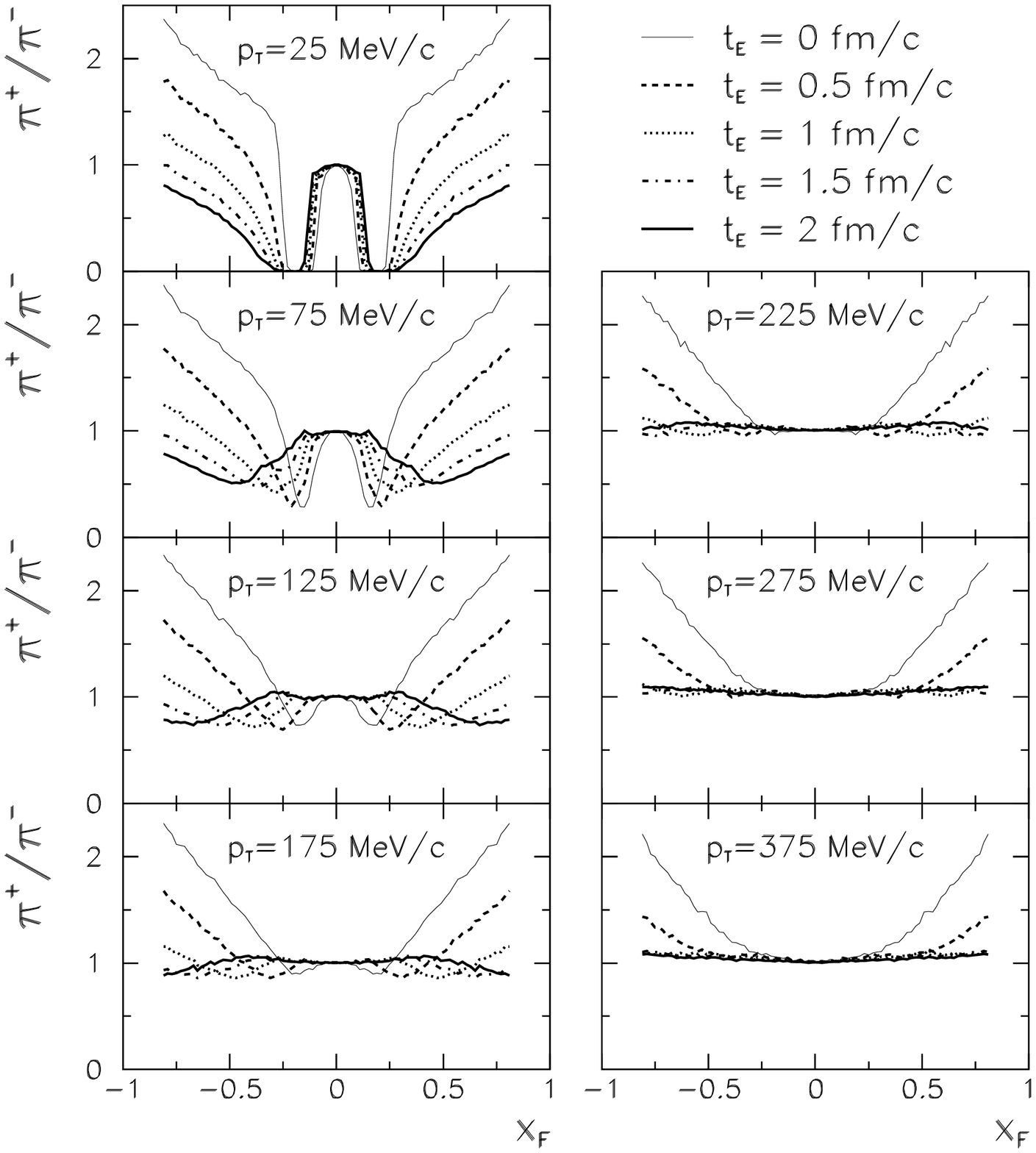}
\end{center}
 \vspace{-0.2cm}
\caption{The $\pi^+/\pi^-$ cross section ratio for the
peripheral Pb+Pb reaction, shown at fixed values of $p_T$ as a function of
$x_F$. The five considered values of the pion emission time $t_E$ are
differentiated by means of different line types. 
}
\label{fig:ratio2}
\end{figure}

In Fig.\ref{fig:ratio2} we show similar ratio as a function of $x_F$
for fixed values of transverse momenta.

\section{Splitting of directed flow of charge pions}

\begin{figure}[t]             
\begin{center}
\includegraphics[width=6cm]{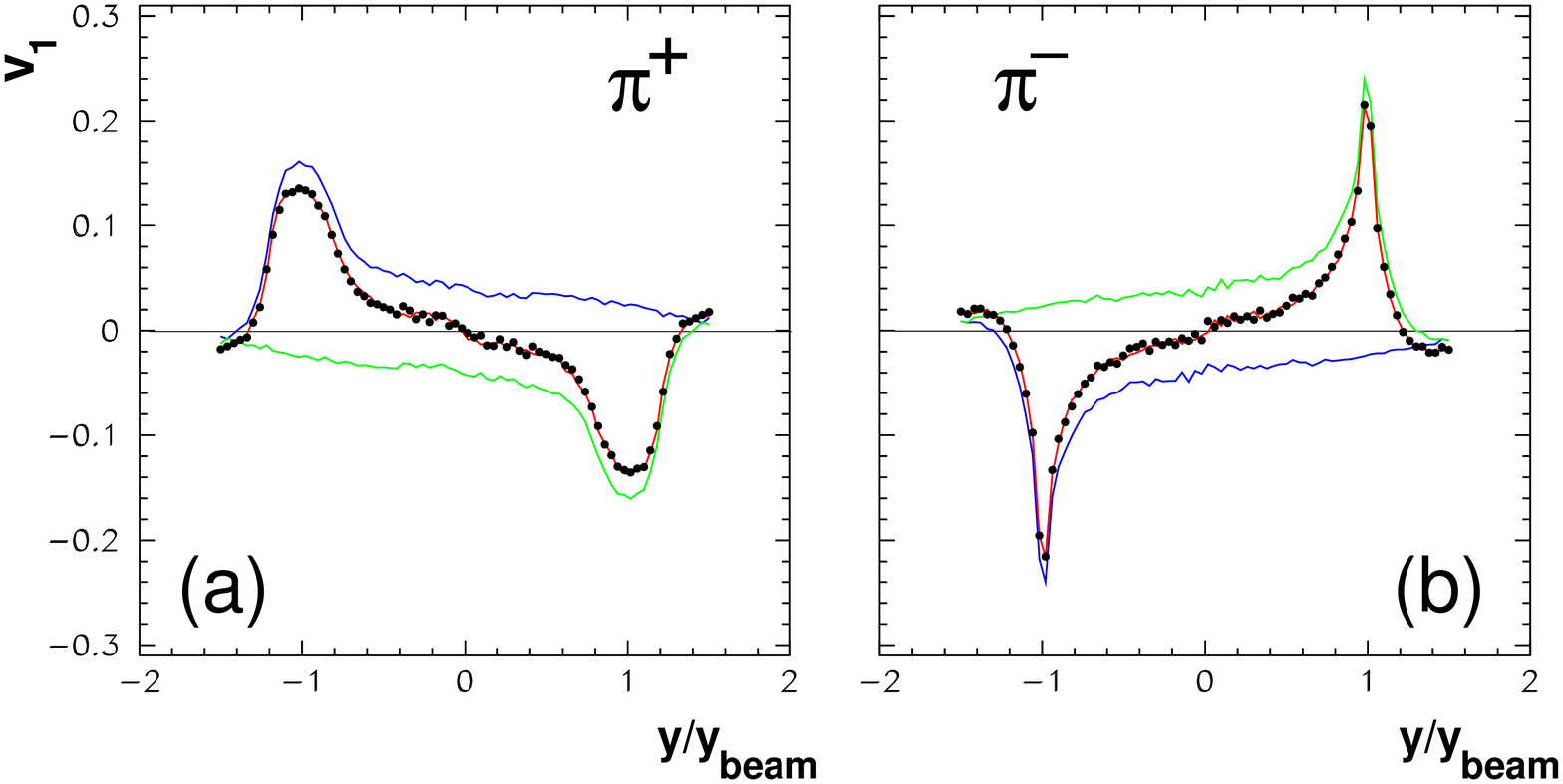}
\end{center}
\caption{(Color online) Spectator-induced electromagnetic effect on 
directed flow of (a) $\pi^+$ and (b) $\pi^-$, in 
peripheral Pb+Pb collisions at $\sqrt{s_{NN}}=17.3$~GeV. The green solid
curve shows the directed flow induced electromagnetically by 
the right (R) spectator. The blue solid curve shows the directed flow 
induced electromagnetically by the left (L) spectator. Black dots show 
the result of the addition of these two curves. The red solid curve
displays the result of the simulation including both spectators. 
Note: all the simulations assume the pion emission time $t_E=0$~fm/c.
}
\label{fig:onespec}
\end{figure}

In Fig.\ref{fig:onespec} we show the comparison of the
EM-induced directed flow 
$v_1$ obtained seperately for each of the two spectators, compared to 
that obtained when both spectators are included in the simulation. 
The figure demonstrates the additivity of these two components in the 
total EM effect.

\begin{figure}[t]             
\centering
\vspace*{-0.2cm}
\includegraphics[width=5cm]{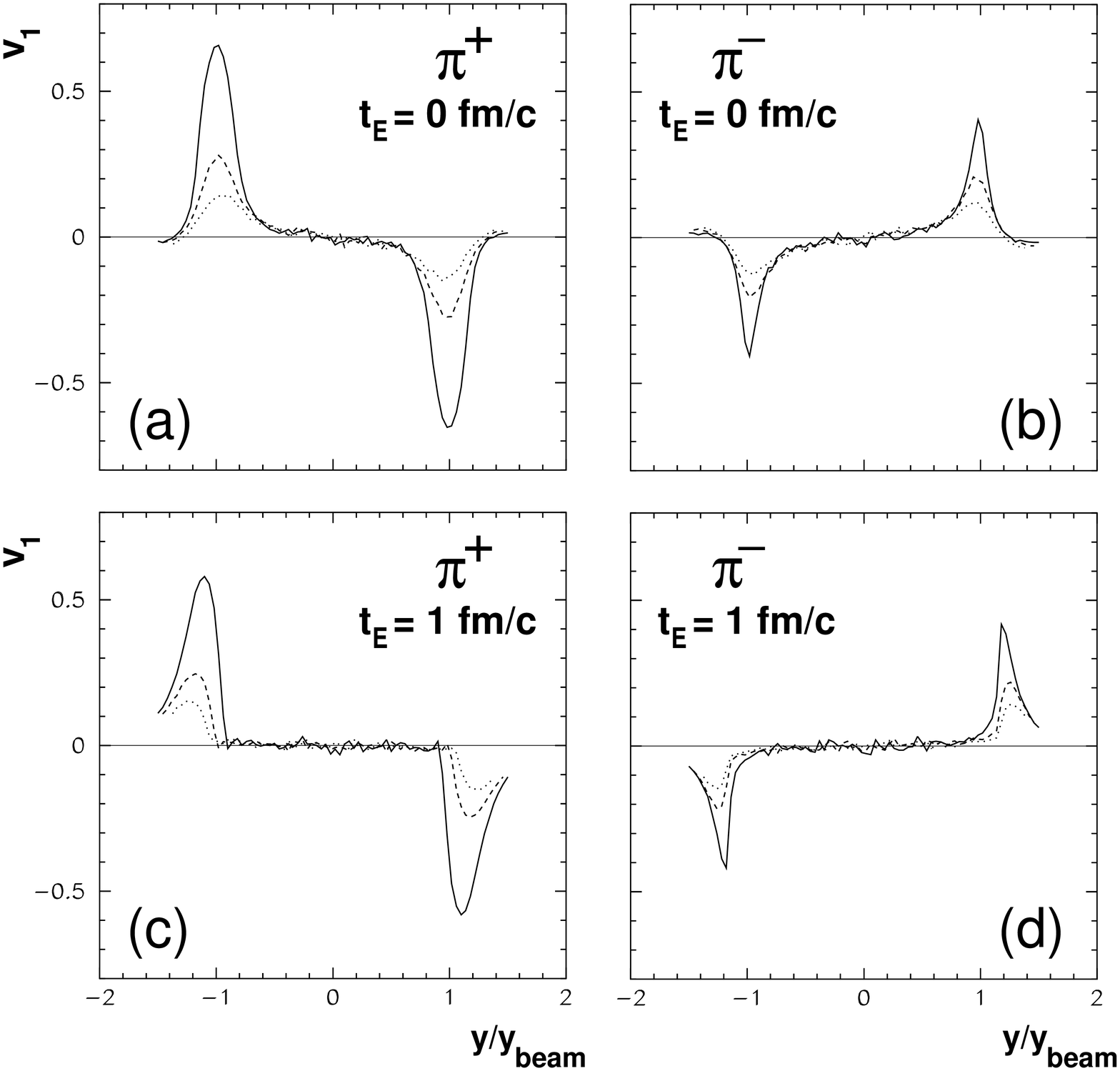}
\caption{Spectator-induced electromagnetic effect on directed flow of 
$\pi^+$ and $\pi^-$ in peripheral Pb+Pb collisions at 
$\sqrt{s_{NN}}=17.3$~GeV, shown at fixed values of pion transverse 
momentum: $p_T=75$~MeV/c (solid), $p_T=125$~MeV/c (dashed),
$p_T=175$~MeV/c (dotted). The top and bottom panels correspond 
to different values of the pion emission time $t_E$.
\vspace*{-0.2cm}}
 \label{fig:pt}
\end{figure}

In Fig.\ref{fig:pt} we show the EM effect on $v_1$ as a function of
rescaled rapidity for two different values of the pion emission time.

\begin{figure}[t]             
\centering
\includegraphics[width=5cm]{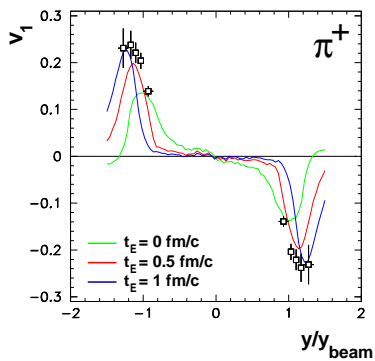}
 \caption{Comparison between experimental data on directed flow of
   positive pions obtained by the WA98 experiment \cite{wa98}), 
and our simulation of electromagnetically-induced directed flow 
of $\pi^+$. The three curves correspond to three different values of 
the assumed pion emission time $t_E$. 
}
 \label{fig:comp}
\end{figure}

Fig.~\ref{fig:comp} shows the experimental data of the WA98 collaboration,
superimposed with the rapidity-dependence of the 
electromagnetically-induced directed
flow as obtained from our work. For the latter, three values of the pion
emission time are assumed: $t_E=0$, $0.5$, and $1$~fm/c. Our results are
integrated over $p_T$ from 0 to 1~GeV/c. 
The curves corresponding to $t_E=0$ 
are the same as in Fig.~\ref{fig:onespec}.

\begin{figure}[t]
\begin{center}
\includegraphics[width=5cm]{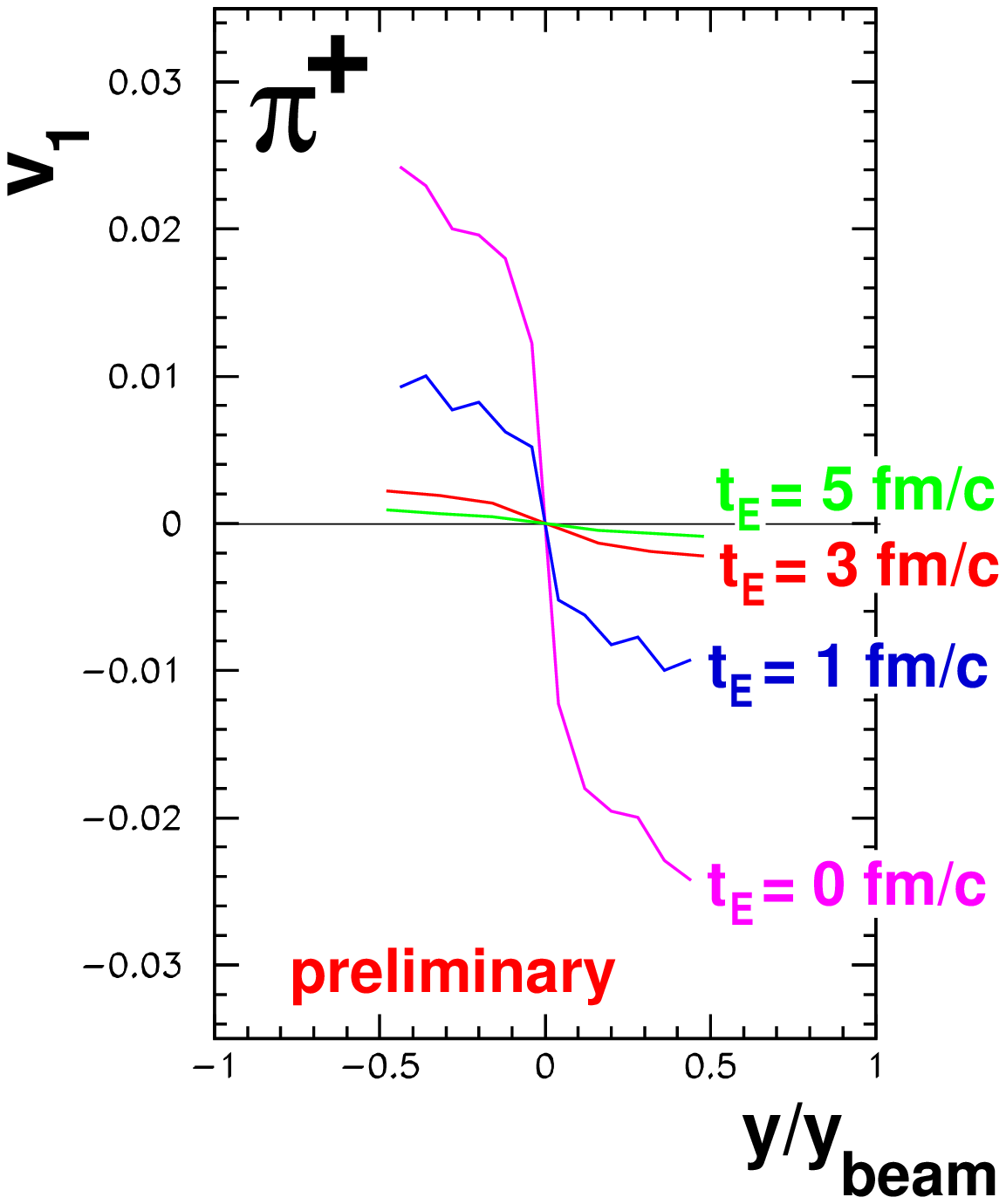}
\includegraphics[width=5cm]{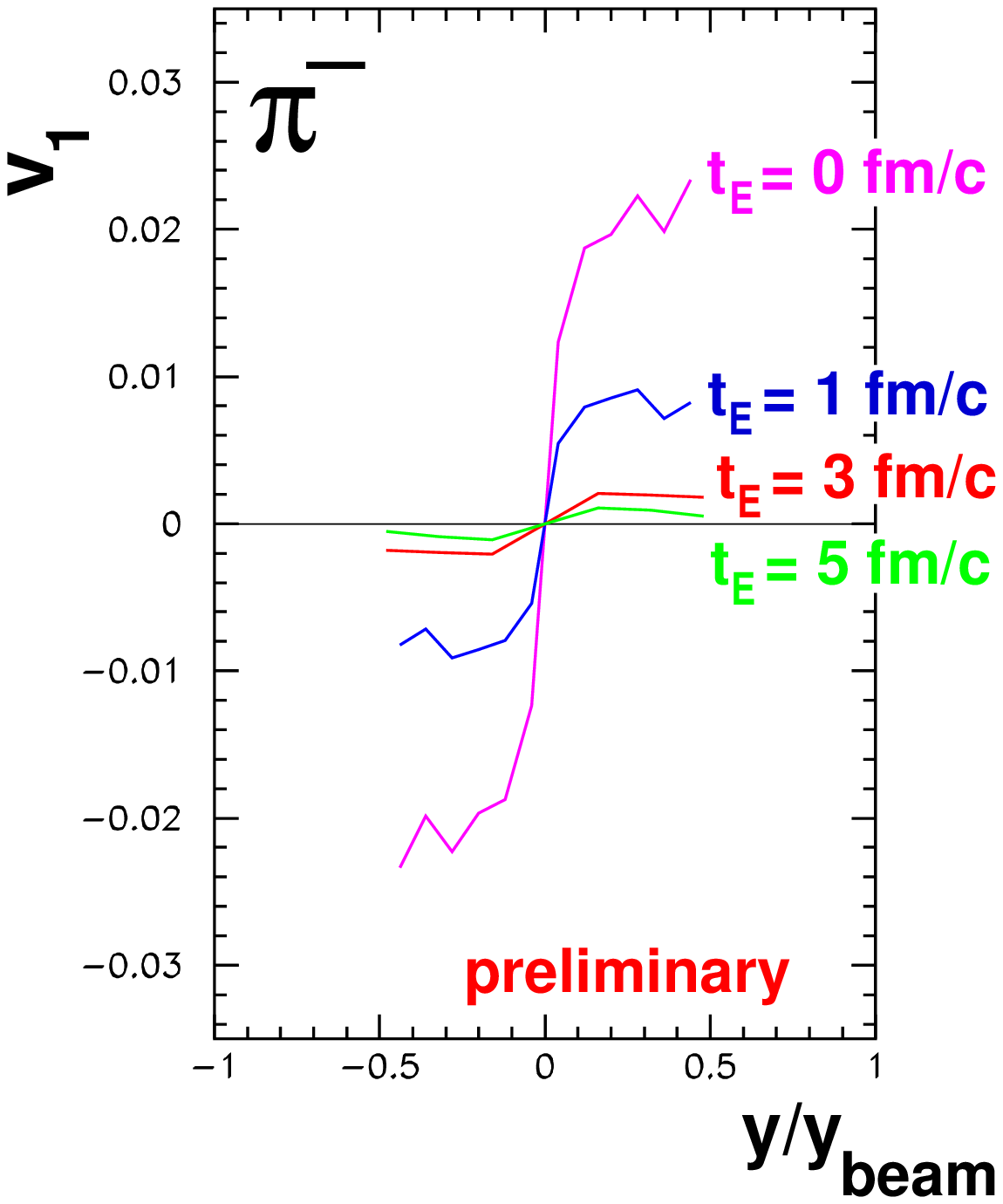}
\caption{Emission time dependence of directed flow.
The simulation has been made for $Au + Au$ collisions at
$\sqrt{s_{NN}}$ = 7.7 GeV.} 
\label{fig:emission_time_dependence}
\end{center}
\end{figure}

In Fig.\ref{fig:emission_time_dependence} we show how $v_1$
depends on the emission time. The shorter emission time the larger
$v_1$. This figure, when compared to experimental data, can
allow to extract the emission time.

\begin{figure}[t]
\begin{center}
\includegraphics[width=5cm]{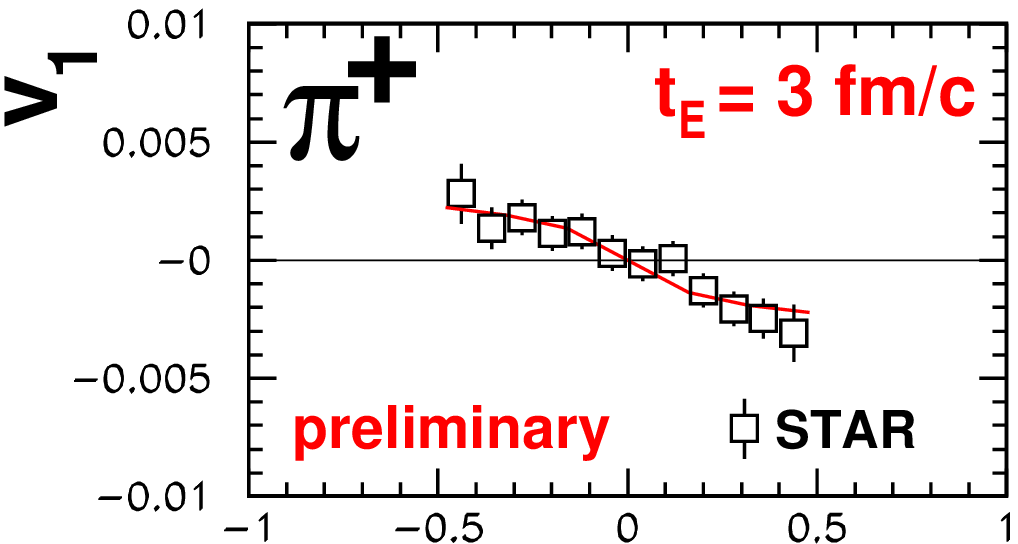}
\includegraphics[width=5cm]{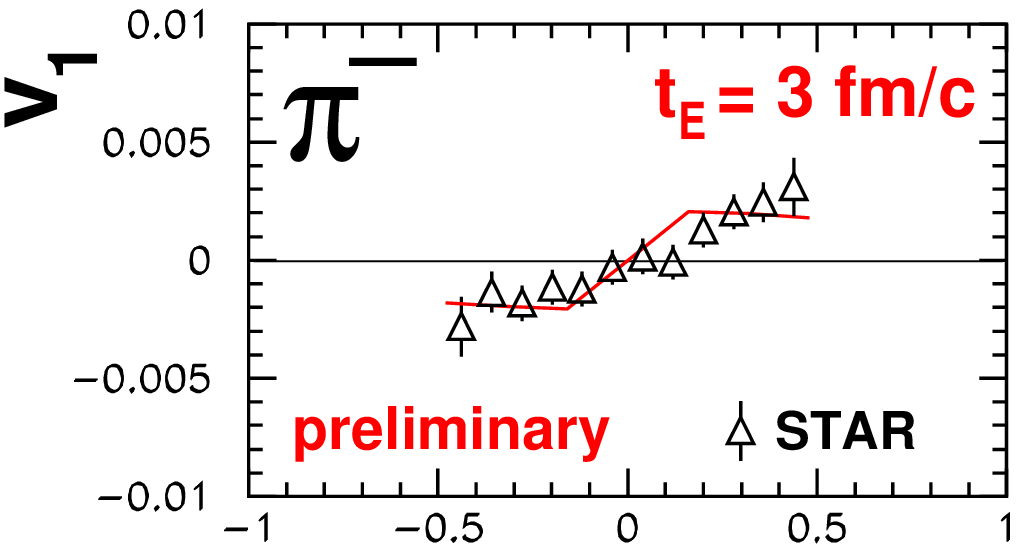}
\caption{Fitting the emission time to STAR data \cite{STAR_data}.}
\label{fig:fitting_emission_time}
\end{center}
\end{figure}

In Fig.\ref{fig:fitting_emission_time} we show the best fit
of the emission time to STAR data \cite{STAR_data}. 
We note that the average directed flow
 $\frac{v_1^{\pi^+}+v_1^{\pi^-}}{2}$
 is subtracted from the STAR data in order to isolate its 
electromagnetic component as discussed above. From this fit we conclude 
that pions are produced at an (average) distance of about 3~fm from the 
spectator system. A more detailed discussion will be given in \cite{RS2014}.

\section{Conclusions and Outlook}

Our investigation concerning electromagnetic effects caused
by fast moving spectators can be summarized as follows:

\begin{itemize}
 \vspace{-0.1cm}
\item The EM interaction caused by the moving remnant charge
produces visible distorsions in the final state
distributions of $\pi^+$ and $\pi^-$.
 \vspace{-0.1cm}
\item The main feature of this electromagnetic effect is a 
 big dip in the $\pi^+$ density distribution 
at low transverse momenta in 
the vicinity of $x_F\approx\pm 0.15$, accompanied by a 
substantial increase of $\pi^-$ density in the same region.
 \vspace{-0.1cm}
\item The effect is clearly sensitive to initial conditions
and carries interesting information on the mechanism of the non-perturbative 
particle production process, and in particular on its evolution in space and time.
 \vspace{-0.1cm}
\item Our study demonstrates the importance of new, 
double-differential data on the $x_F$ and $p_T$-dependence 
of pion production in peripheral nucleus+nucleus collisions.
 \vspace{-0.1cm}
\item
Electromagnetic fields generated by charged, fast spectators lead
to extra azimuthal distortions and contribute to directed flow.
 \vspace{-0.1cm}
\item
The above effect is opposite for $\pi^+$ and $\pi^-$ 
and leads to a splitting of $v_1$. 
This splitting is superimposed on other effects (like hydrodynamics).
 \vspace{-0.1cm}
\item
This effect is confirmed by the WA98 and STAR data.
 \vspace{-0.1cm}
\item
The splitting strongly depends on the emission time of pions
and can be therefore used to measure the emission time.
 \vspace{-0.1cm}
\item
The splitting depends on the transverse momentum
of pions. 
\end{itemize}

The topic discussed here requires further investigations.
In particular:

\begin{itemize}
 \vspace{-0.1cm}
\item Precise data for $\pi^+$ and $\pi^-$ and different energies are needed.

 \vspace{-0.1cm}
\item The dependence on rapidity and transverse momentum should be analyzed.

 \vspace{-0.1cm}
\item Realistic modelling of the source is badly needed.

 \vspace{-0.1cm}
\item A procedure to extract emission time would be very 
      useful and provide a complementary information to, e.g., HBT measurements.

 \vspace{-0.1cm}
\item The evolution time of spectator systems should be 
      better understood.

 \vspace{-0.1cm}
\item Other harmonics are also subjected to spectator EM splitting. 

 \vspace{-0.1cm}
\item Exploration of the above electromagnetic effects as a function of
 particle species, rapidity,
      transverse momentum and centrality will provide valuable insight into the 
      non-perturbative dynamics of the heavy ion collision.
\end{itemize}

{\bf Acknowledgments}\\

This work was supported by the Polish National Science Centre 
(on the basis of decision no. DEC-2011/03/B/ST2/02634).


\end{document}